\pdfoutput=1
\documentclass[a4paper,11pt]{article}
\usepackage{amsmath}
\usepackage{graphicx}
\usepackage{subfigure}
\usepackage[T1]{fontenc} 
\usepackage{accents} 

\newcommand{\set}[1]{\mathbf{#1}} 
\newcommand{\abar}{\bar{a}}

\title{Reciprocity of mobile phone calls} \author{Lauri Kovanen, Jari
  Saram\"{a}ki, Kimmo Kaski\\{\small Department of Biomedical
    Engineering and Computational Science}\\{\small Aalto
    University School of Science and Technology, Finland}\\{\small
    lauri.kovanen@tkk.fi}} \date{}

\begin{document}

\maketitle

\begin{abstract}
We present a study of the reciprocity of human behaviour based on
mobile phone usage records. The underlying question is whether human
relationships are mutual, in the sense that both are equally active in
keeping up the relationship, or is it on the contrary typical that
relationships are lopsided, with one party being significantly more
active than the other. We study this question with the help of a
mobile phone data set consisting of all mobile phone calls between 5.3
million customers of a single mobile phone operator. It turns out that
lopsided relations are indeed quite common, to the extent that the
variation cannot be explained by simple random deviations or by
variations in personal activity. We also show that there is no
non-trivial correlation between reciprocity and local network density.
\end{abstract}

\section{Introduction}

Representing collections of social relations as networks has become a
standard practice not only in sociology but also in the study of
complex systems. A social network $G = (V,E)$ is defined by the set of
$N$ nodes, $V = \{v_i\}_{i=1}^N$, which correspond to individuals, and
by the set of edges (or links) $E = \{e_{ij}=(v_i,v_j),\,v_i,v_j \in
V\}$, which correspond to some associations between the
individuals. Even though it is common to assume the associations to be
symmetric (i.e. $e_{ij} = e_{ji}$) \cite{Goodreau2007,
  Onnela_PNAS_2007, Liljeros2001}, for our purposes the edges are
always directed, meaning that $e_{ij}$ is independent of $e_{ji}$. In
a \emph{weighted} network each edge is also assigned a weight. We
consider the edge weights as proxies for the strength of a
relationship, and therefore the weights are strictly positive. In our
notation $w_{ij} > 0$ is the weight of a directed edge from node $v_i$
to $v_j$, and $w_{ij} = 0$ is equivalent to saying that there is no
edge from $v_i$ to $v_j$.

We use the term \emph{reciprocity} to depict the degree of mutuality
of a relationship. A relationship with a high reciprocity is one where
both are equally interested in keeping up the relationship --- a good
example is the relationship of Romeo and Juliet in the famous play
with the same name by William Shakespeare, where the two lovers strive
to share each other's company despite their families' objections. On
the other hand, in a relationship with a low reciprocity one person is
significantly more active in maintaining the relationship than the
other. We judge reciprocity from actual communications taking place
between people.\footnote{Of course, the reciprocity deduced from
  communication, even when the communication is completely known, does
  not necessarily match the individuals' \emph{perceptions} of the
  reciprocity.} In Shakespeare's time this would have meant counting
the number of poems passed and sonnets sung, but in our modern era it
is easier to make use of the prevalence of electronic communication.

From the sociological point of view the directionality of associations
is interesting \emph{per se}: are human relationships typically mutual
or not? If not, what possible explanations are there for the lopsided
relations? Directionality of communication might also affect for
example the spreading of mobile viruses \cite{Wang2009}, or the
spreading of information and ideas in societies. There is also
evidence that the reciprocity of a relationship is a good predictor of
its persistence \cite{HidalgoPhysA2008}.

We note that even though symmetrical associations are common in the
social networks literature, this is rarely meant as a claim of perfect
reciprocity. For many purposes the assumption of undirected edges is a
useful simplification or it follows directly from the definition of
association. This is the case for example with scientific
collaboration networks \cite{Newman2001}, where we draw a link between
two scientists if they have collaborated in writing an article. On the
other hand, some associations are intrinsically directed, like e-mail
networks \cite{Holme2005} or friendship networks. The latter was shown
explicitly in \cite{Goodreau2007} with data from the National
Longitudinal Study of Adolescent Health, where US students were asked
to name up to 5 of their best friends: only 35 \% of the roughly 7000
such nominations were reciprocated. Moreover, some undirected networks
can be considered as directed networks when examined more closely. For
example, collaboration networks could be extended by adding
information on which party initiated the collaboration; the same is
true for sexual networks \cite{Liljeros2001}.

Previous studies have mostly considered reciprocity as a global
feature of a directed, unweighted network. They are based on the
classical definition according to \cite{Wasserman1994} that quantifies
reciprocity as $r = \frac{L^{\leftrightarrow}}{L}$ where $L = \sum_{i
  \neq j}a_{ij}$ is the total number of (directed) links in the
network and $L^{\leftrightarrow} = \sum_{i \neq j} a_{ij}a_{ji}$ is
the number of the mutual edges (the latter sum goes over all pairs of
nodes). Here $a_{ij}$ are the elements of the adjacency matrix such
that $a_{ij}=1$ if there is a directed link from node $v_i$ to $v_j$,
and $a_{ij} = 0$ otherwise. Defined this way, $r = 0$ if there are no
bidirectional links, and $r=1$ when all links in the network are
bidirectional. For example \cite{Newman2002} finds that one network
constructed from email address books has $r = 0.231$.

One problem with this definition of reciprocity is that it depends on
the density of the network. A large value of $r$ is more significant
when encountered in a sparse network, while a small $r$ is a
surprising finding in a dense network. The value of $r$ must thus be
compared to the network density $\abar = \frac{L}{N(N-1)}$, which also
equals the reciprocity of a fully random
network. In \cite{Garlaschelli2004} this is taken further by combining $r$
and $\abar$ into a single measure of reciprocity, $\rho =
\frac{r-\abar}{1-\abar}$. This idea is extended in
\cite{Zamora-Lopez2008} by comparing $r$ in the original network to
that in randomized networks when the degree sequence or various degree
correlations are held constant. It turns out that in most networks
this is enough to explain most of the reciprocity.

The unweighted treatment of reciprocity however has a severe problem
when dealing with social networks: a truly unidirectional social
contact is an extremely rare thing to find in everyday social
environment. When such edges do occur, for example in the
above-mentioned study where the subjects were asked to name a fixed
number of acquaintances \cite{Goodreau2007}, the unidirectional links
are mostly artifacts of the research method. They can however be
interpreted as a sign of an underlying edge bias, a cue that the
relationship is not seen alike by the two people involved.

Unlike in the unweighted case, edge weights allow us to study
reciprocity as a property of a single edge instead of the full
network. To this end we define the \emph{edge bias} between nodes
$v_i$ and $v_j$ as
\[
b_{ij} = \frac{w_{ij}}{w_{ij}+w_{ji}}.
\]
The edge bias is the simplest possible measure for studying
reciprocity. Note that because $b_{ij} + b_{ji} = 1$, the distribution
of $b_{ij}$ for all edges in the network is symmetric around $0.5$ and
therefore it suffices to study only the range $b_{ij} \in [0.5,\,1]$.

\section{Data}

We study the reciprocity of communication in a mobile phone data set
consisting of 350 million calls between 5.3 million customers made
during a period of 18 weeks. Mobile phone calls provide an excellent
proxy for studying this kind of reciprocity because phone calls are at
the same time directed and undirected: The calls are directed because
it is the caller who decides to make the call and invest his time (and
often money) in that particular relationship, but since the
conversation during the call is undirected, there is no immediate
reason for the recipient to call back shortly after. Consider the
difference to SMS messages or emails where a conversation means
sending a sequence of reciprocated messages, making it more difficult
to study reciprocity by simply counting the number of messages.

To remove possibly spurious contacts, i.e. those that are more likely
to be chance encounters instead of true social relationships, we have
preprocessed the data by removing all unidirectional edges. Thus the
unweighted reciprocity according to the definition above would be
$r=1$, and therefore the methods for analyzing the reciprocity of
unweighted, directed networks are not applicable.\footnote{To be
  exact, $r \approx 1$ because any communication, either SMS message
  or call, was sufficient to make an edge bidirectional, and thus
  there are some edges where only one person has made calls.} The
weight $w_{ij}$ is the total number of calls customer $v_i$ made to
$v_j$ during the whole 18 week period.

\begin{figure} \centering
\includegraphics[width=7cm]{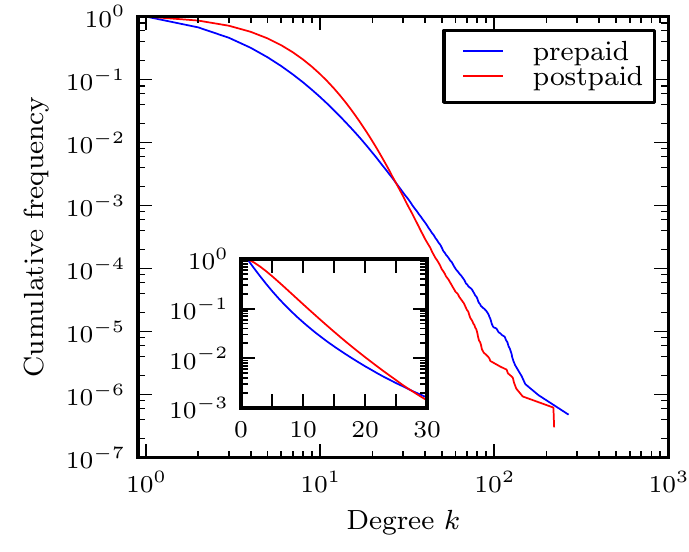}
\caption{Total degree distribution of the mobile phone data set
  separately for prepaid and postpaid users. Inset: The distribution
  until $k=30$ in semi-logarithmic coordinates.}
\label{fig:degree_distribution}
\end{figure}

The customers can be split into two groups based on how they pay for
their mobile phone service: \emph{prepaid} users pay for their usage
beforehand, \emph{postpaid} users pay afterword. While this initial
difference might seem small, it results in major differences in the
basic statistics and it is therefore necessary to analyze the two user
groups separately. The statistics of calls between prepaid and
postpaid users would mostly just reflect the differences in degree and
strength distributions (discussed below). Thus in this article we only
study calls between pairs of prepaid users and calls between pairs of
postpaid users. The two user groups are approximately equal in size.

The basic statistics of the network often imprint also more complex
statistics, and it is therefore a good idea to first go through some
basic distributions. The \emph{degree} $k_i$ of a node is defined as
the number of neighboring nodes. Because all edges in our network are
bidirectional after preprocessing, the out-degree (number of people
called) always equals the in-degree (number of people from whom a call
has been received). The degree distributions of the network, shown in
Figure \ref{fig:degree_distribution} separately for prepaid and
postpaid users, have long power-law tails. The average degree was
found to be 3.41 for prepaid users and 5.15 for postpaid users.

\begin{figure} \centering
\subfigure[]{
  \includegraphics[width=7cm]{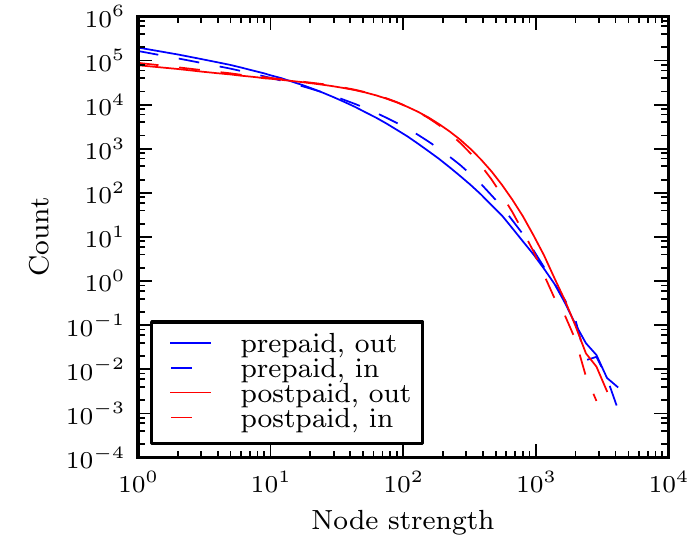}
  \label{fig:strength_distribution}
}
\subfigure[]{
  \includegraphics[width=7cm]{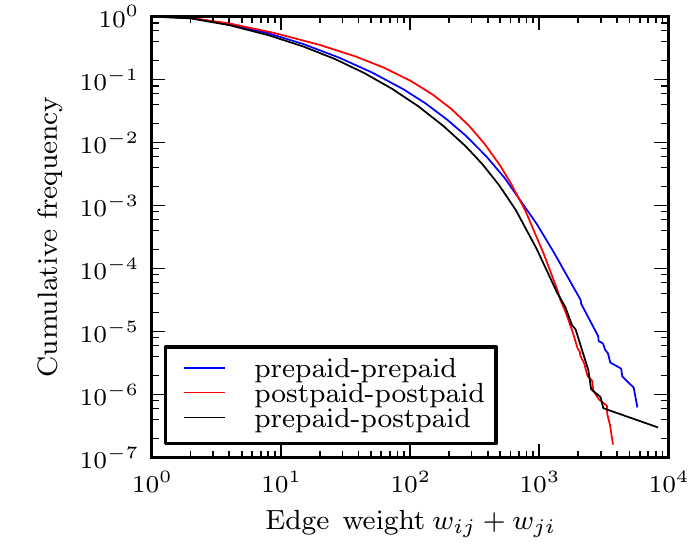}
  \label{fig:weight_distribution}
}
\caption{\textbf{(a)} The in- and out-strength distributions for the two users
  types. \textbf{(b)} The total weight distribution for edges between
  the two user types.}
\end{figure}

The \emph{strength} of a node is defined as the sum of weights of
adjacent edges. For a directed network we can define both the
out-strength $s_i^{\textrm{out}} = \sum_j w_{ij}$, which is the total
number of calls made, and the in-strength $s_i^{\textrm{in}} = \sum_j
w_{ji}$, the total number of calls received. The strength
distributions in Fig. \ref{fig:strength_distribution} turned out also
to be fat-tailed, and we can see that postpaid users make much
more calls. In fact, postpaid users make on average 10 times more
calls than prepaid users. We can also see that prepaid users receive
more calls than they make, while the most active postpaid users make
more calls than they receive.

Figure \ref{fig:weight_distribution} shows the total edge weight
distribution for edges between users of the same or different
type. There are more calls between users of the same type, which
suggests that the user type is correlated between acquaintances.

The data set only includes the customers of a single mobile phone
operator that has about 20 \% market share in the country in
question. While there are calls from and to people outside this
customer base, this has no skewing effect on our study because our
analysis concentrates on communication between the known customers,
and this information is completely known.


\section{Results}
\label{sec:results}

\begin{figure} \centering
\subfigure[Prepaid]{
\includegraphics[width=7cm]{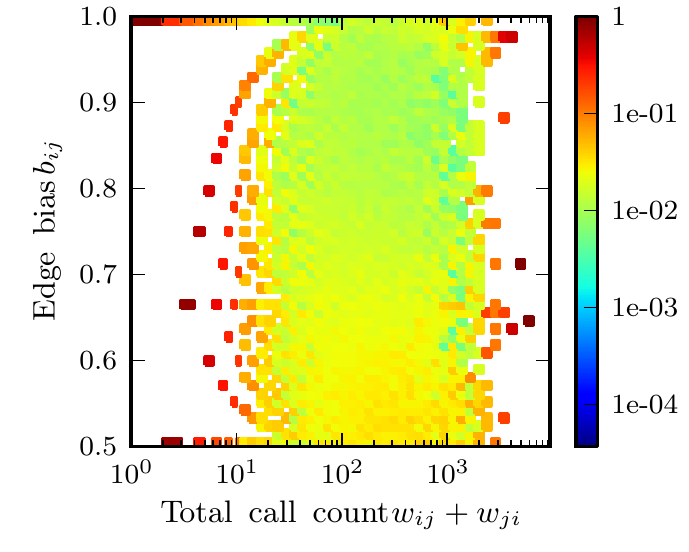}
\label{fig:bias_dist_prepaid}
} \subfigure[Postpaid]{
\includegraphics[width=7cm]{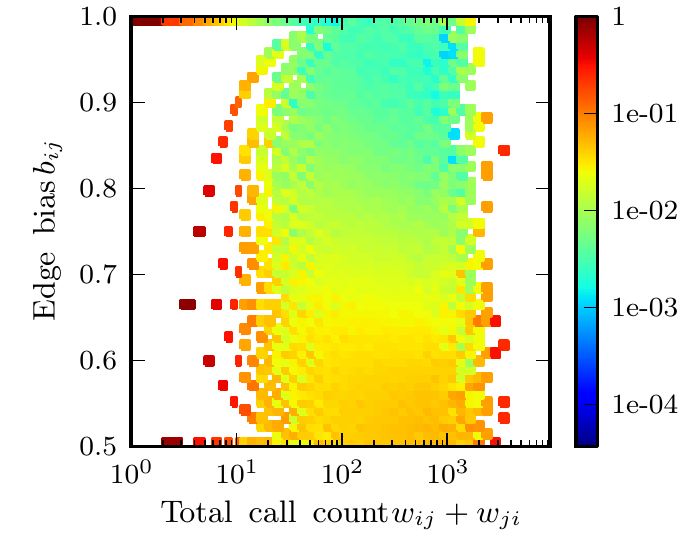}
\label{fig:bias_dist_postpaid}
}
\caption{The distribution of edge bias values as a function of edge
  weight for \textbf{(a)} prepaid and \textbf{(b)} postpaid
  users. Each vertical strip in the figure is a distribution and sums
  up to one. When $w_{ij} + w_{ji}$ is small, we can see that the edge
  bias is strongly quantized; for example if the total weight is 5,
  there are only three possible values for the edge bias in the range
  $b_{ij} \in [0.5,\,1]$: $\frac{3}{5} = 0.6$, $\frac{4}{5} = 0.8$ or
  $\frac{5}{5} = 1.0$. The white area on right on the other hand is
  caused by missing values; there are not many edges with over 2000
  calls in 18 weeks.}
\end{figure}

\begin{figure} \centering
\subfigure[]{
  \includegraphics[width=7cm]{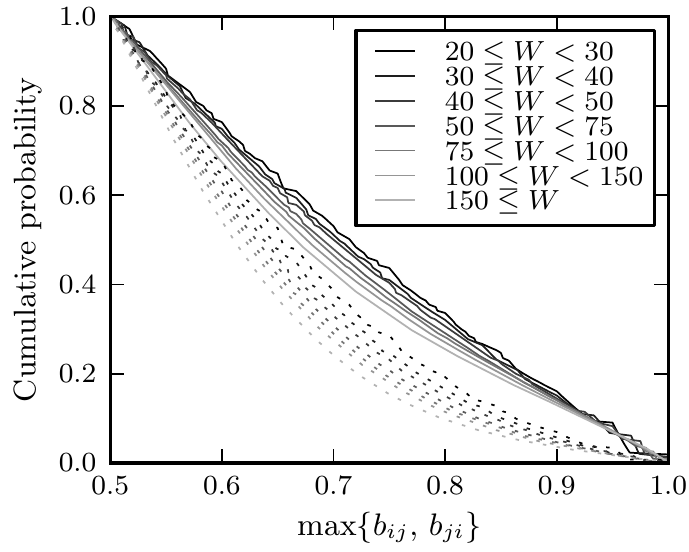}
  \label{fig:cumulative_bias_by_weight}
}
\subfigure[]{
  \includegraphics[width=7cm]{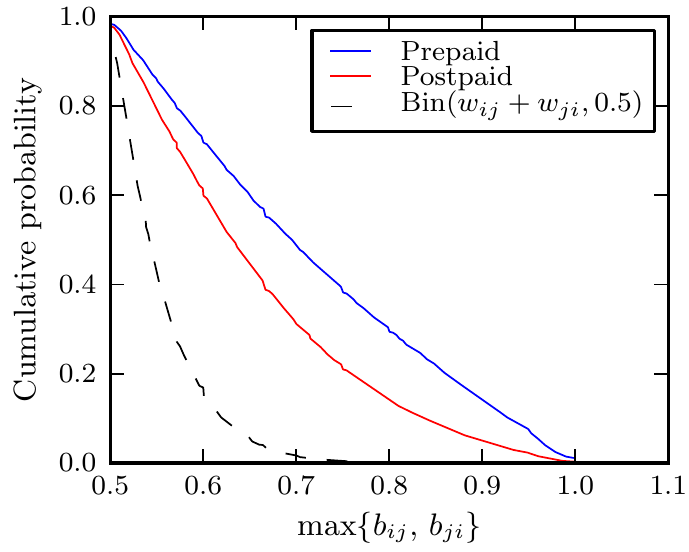}
  \label{fig:binomial_comparison}
} 
\subfigure[]{
  \includegraphics[width=7cm]{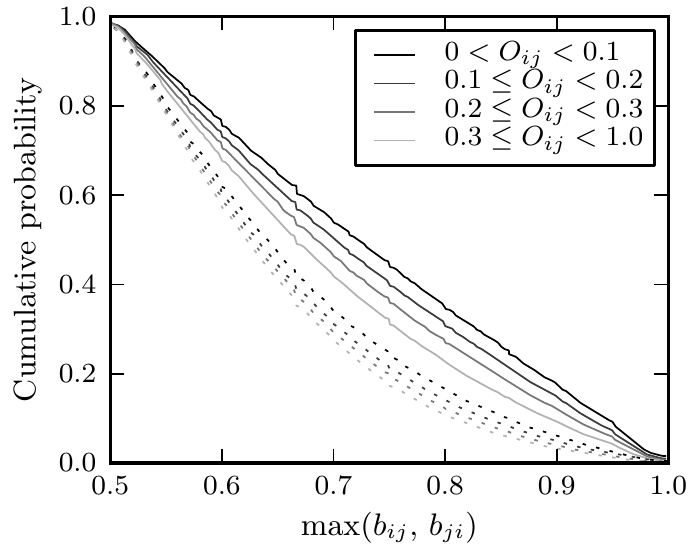}
  \label{fig:cumulative_bias_by_overlap}
}
\subfigure[]{
  \includegraphics[width=7cm]{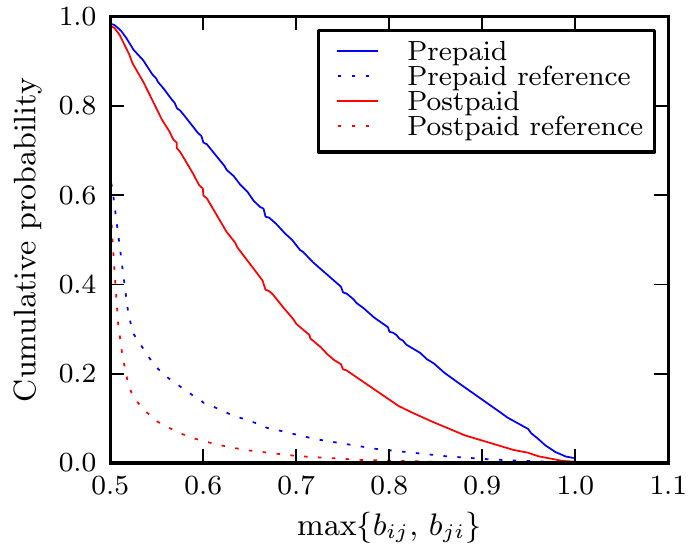}
  \label{fig:reference_comparison}
}
  \caption{\textbf{(a)} The cumulative edge bias distribution
    separately for different ranges of total edge weight. The solid
    lines are for prepaid users and the dotted lines for postpaid
    users, lighter color means higher edge weight. \textbf{(b)} The
    cumulative edge bias distribution for prepaid and postpaid users
    compared to the cumulative binomial distribution. In theory the
    binomial reference is slightly different for prepaid and postpaid
    users due to different edge weight distributions, but in practice
    the difference is negligible. \textbf{(c)} The cumulative edge
    bias distribution separately for different ranges of overlap. The
    solid lines are for prepaid users and the dotted lines for
    postpaid users, lighter color means higher overlap. It would seem
    that larger overlap implies edge bias values more concentrated
    around 0.5, but as shown in Figure
    \ref{fig:bias_by_weight_and_overlap}, this is a spurious
    correlation. \textbf{(d)} The cumulative edge bias distribution
    for prepaid and postpaid users compared to the references where
    the out-strength of each node has been redistributed in an attempt
    to equalize edges. In all plots only edges with $w_{ij} + w_{ji}
    \geq 20$ are included to avoid excessive quantization of the
    cumulative distribution. With 18 weeks of data this requirement
    translates to an average of roughly one call (in either direction)
    per week. These edges contain about 85 \% of all calls in the
    network.}
\end{figure}

Figures \ref{fig:bias_dist_prepaid} and \ref{fig:bias_dist_postpaid}
show the distribution of $b_{ij}$ for prepaid and postpaid users as a
function of total edge weight. We can see that even though values
around 0.5 are more common, there are still many edges with larger
values of $b_{ij}$. This becomes more obvious in Figure
\ref{fig:cumulative_bias_by_weight} that shows the cumulative
distribution of edge bias for different weight ranges. For example
among prepaid users the edges with $b_{ij} \geq 0.8$, i.e. edges where
one participant makes more than 80 \% of all calls, make up over 25 \%
of all the edges even when the edge has over 100 calls. For postpaid
users the numbers are somewhat smaller.

Naturally, we should not expect to find \emph{perfect} reciprocity on
all edges, but to justify treating social networks as undirected there
should be some tendency towards $b_{ij}=0.5$. But how much? The
simplest step away from perfectly reciprocal relationships is to
assume each dyad to be reciprocal in the probabilistic sense: both
participants are equally likely to make a call. Thus, even though
edges with only few calls could still be very biased, edges with 50 or
more calls should have $b_{ij}$ very close to $0.5$. This idea is also
supported by Figure \ref{fig:cumulative_bias_by_weight}, where the
distributions of high weight edges are somewhat more concentrated
around $0.5$.

The hypothesis about equal likelihood for making a call implies that
the edge bias values should follow a binomial distribution with
$p=0.5$ and $n=w_{ij}+w_{ji}$. However, Figure
\ref{fig:binomial_comparison} shows that such a binomial distribution
is far from truth: if the calls were evenly distributed between both
participants, there should be almost no edges with $b_{ij} > 0.7$.

But is it reasonable to expect the edge bias values to be binomially
distributed in the first place? As shown in
Fig. \ref{fig:strength_distribution}, the node strengths vary widely,
which itself might be enough to force large deviations from $b_{ij} =
0.5$ --- in other words the relationships would only appear biased
because some people tend to call more than others. To test this claim
we redistribute the total out-strength of each node onto its already
existing edges so as to make the edge bias values as close to 0.5 as
possible (see Appendix for details). It turns out that it is in fact
possible to significantly reduce the number of high edge bias values,
especially among postpaid users, as shown in
Fig. \ref{fig:reference_comparison}. This proves that the strength
distribution is not a sufficient explanation for the large edge bias
values.

Since it appears that the edge bias cannot be explained away by
properties of the node but is instead an intrinsic property of the
edge itself, it is interesting to see whether it correlates with other
local features of the network. Is there for example a connection
between the edge bias and the community structure of the network? We
measure local community structure with \emph{edge overlap}, defined as
$O_{ij} = n_{ij}/((k_i- 1) + (k_j - 1) - n_{ij})$, where $n_{ij}$ is
the number of common neighbors of nodes $v_i$ and $v_j$. Thus the edge
overlap gives the fraction of common neighbors out of all possible
common neighbors, and is therefore larger in denser parts of the
network.

\begin{figure} \centering
\includegraphics[width=7cm]{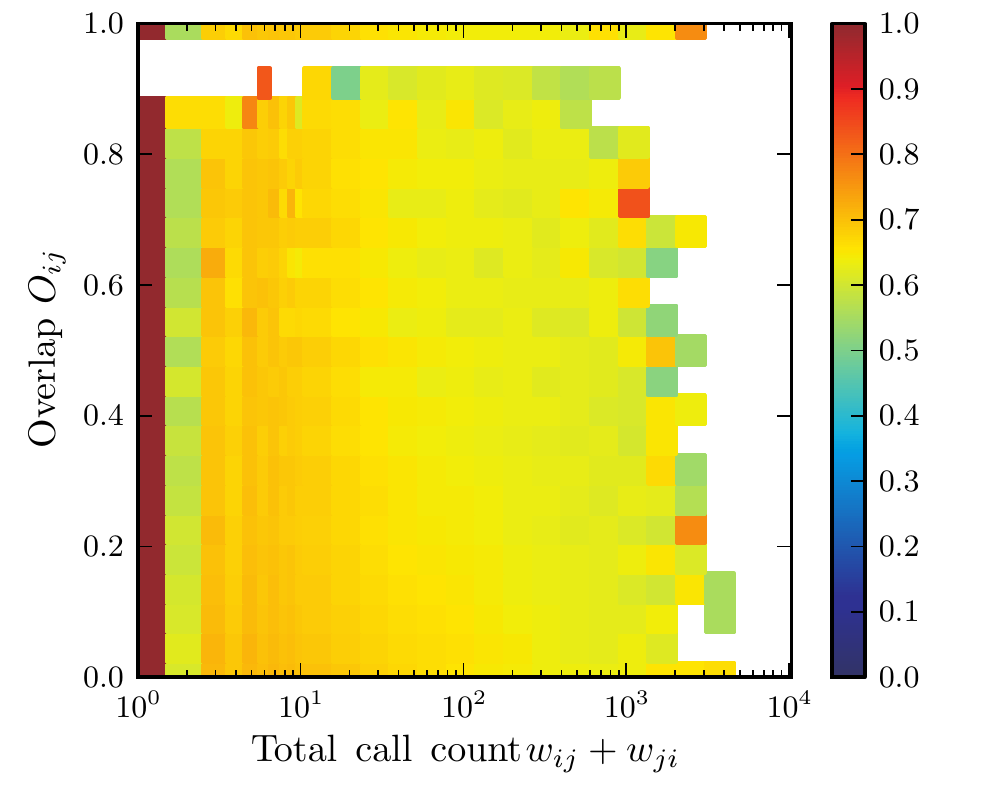}
\caption{The average edge bias as a function of the edge weight and
  overlap for postpaid users (the corresponding plot for prepaid users
  is qualitatively similar). The color tells the average edge bias for
  edges with given total weight and edge overlap. The average edge
  bias changes with the edge weight (moving left to right) but does
  not significantly change with the edge overlap (moving down to up)
  given the edge weight.}
\label{fig:bias_by_weight_and_overlap}
\end{figure}

Looking at Figure \ref{fig:cumulative_bias_by_overlap} it seems that
the edge bias does indeed correlate with overlap: larger overlap
implies an edge bias more concentrated around 0.5 for both prepaid and
postpaid users. However, Figure \ref{fig:bias_by_weight_and_overlap}
shows that the observed effect is in fact a spurious result of two
other correlations. The first correlation is the one between edge bias
and weight, shown in Fig. \ref{fig:cumulative_bias_by_weight}. The
second correlation is the \emph{Granovetter hypothesis}
\cite{GranovetterWeakTies1973}, which states that edges between
communities should be weaker than those inside communities. It was
shown in \cite{Onnela_PNAS_2007} that in a mobile phone network this
is indeed the case: edge weights are higher in denser parts of the
network.

\begin{figure} \centering
\subfigure[Prepaid]{
  \includegraphics[width=7cm]{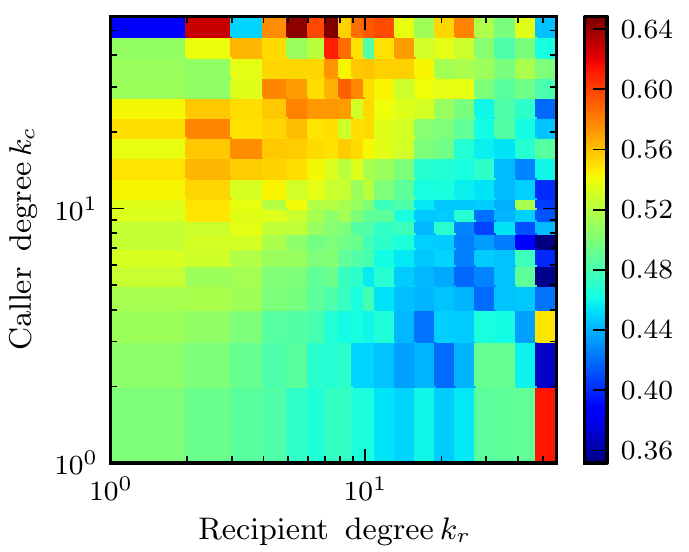}
  \label{fig:degdeg_bias_call_count_prepaid}
}
\subfigure[Prepaid]{
\includegraphics[width=7cm]{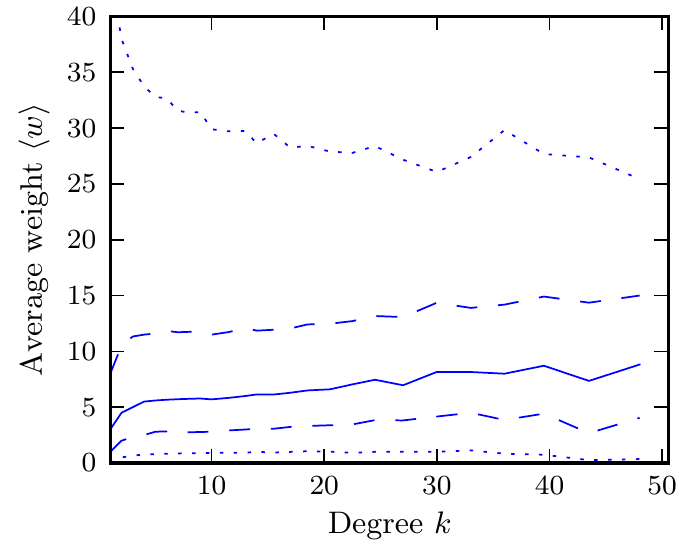}
\label{fig:avg_out_weight_prepaid}
} \\
\subfigure[Postpaid]{
  \includegraphics[width=7cm]{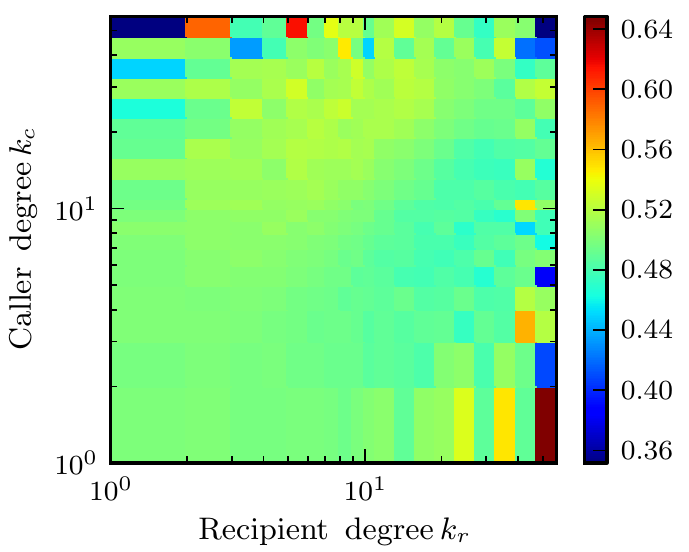}
  \label{fig:degdeg_bias_call_count_postpaid}
}
\subfigure[Postpaid]{
  \includegraphics[width=7cm]{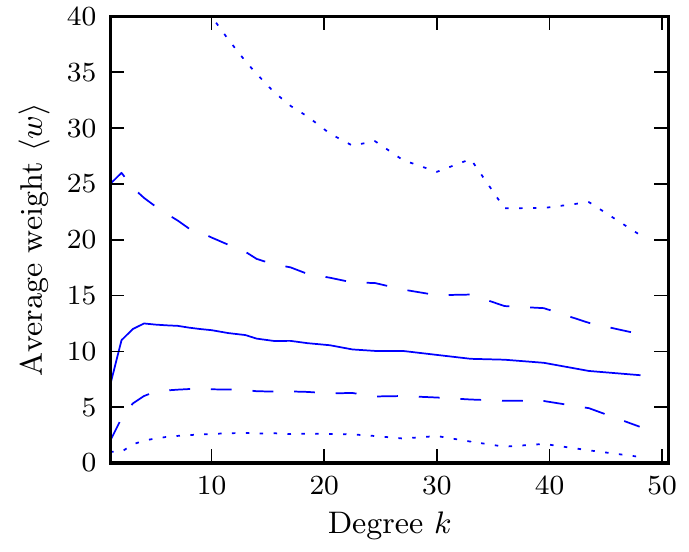}
  \label{fig:avg_out_weight_postpaid}
}
\caption{\textbf{(a,c)} The (weighted) average edge bias as function
  of caller degree $k_c$ and recipient degree $k_r$ for (a) prepaid
  and (c) postpaid users. The color tells the weighted average of the
  edge bias values have been weighted by total edge weight. The
  weighted average puts more emphasis on more active edges and is
  therefore a better indicator for the mass of calls taking place in
  the network. \textbf{(b,d)} The average out-weight as a function of
  node degree for (b) prepaid and (d) postpaid users. The solid line
  is the median, dashed lines give the 25th and 75th and the dotted
  lines the 5th and 95th percentiles. One can see from (a) and (c)
  that on average the greater degree participant is more active. While
  this effect is much more subtle for postpaid users, it is partly
  because it goes against the average edge weight shown in (d).}
\end{figure}

Finally, Figures \ref{fig:degdeg_bias_call_count_prepaid} and
\ref{fig:degdeg_bias_call_count_postpaid} show that there is a
connection between the edge bias and the degrees of the caller and the
recipient. This simply confirms the everyday observation that people
with many contacts tend to be active in keeping up those contacts, as
opposed to simply waiting for the others to call them. While for prepaid
users this observation results at least partly from the fact that
average out-weight increases with degree
(Fig. \ref{fig:avg_out_weight_prepaid}), this is not the case with
postpaid users. In fact, of the edges between two postpaid users the one
with a larger degree is on average more active, even though the
average edge weight decreases with degree
(Fig. \ref{fig:avg_out_weight_postpaid}).

\section{Discussion}

We have shown that highly biased edges where one participant is much
more active than the other are abundant in a mobile phone
network. Although the strength distribution can explain some of the
observed high bias values, it is by no means a sufficient explanation.

Making a phone call is always an investment in a relationship: not
only does it cost time and money to the caller, but the caller must
also make a conscious decision to make the call. While people are in
general known to strive to make their relationships reciprocal, we can
observe plenty of relationships with a striking lack of reciprocity.

It was mentioned earlier that the fact that the data only includes 20
\% of the population has no adverse effect on the data. This should
now be quite obvious, since we have only looked at the edges of which
we have full information. However, we can state something even
stronger: it is likely that any additional data would only
\emph{improve} the reference calculated above, because there would be
more edges to choose from during the optimization. This of course
assumes that the edge bias on the missing edges is distributed
similarly as in the current data, which we feel is a very reasonable
assumption.


We have so far deferred the discussion about whether the results can
be extended to human relations in general. There are obviously many
factors that influence the reciprocity of a relationship, as is
evident from the differences between prepaid and postpaid users. We
note that these difference do not necessarily stem straight from the
differences in paying. For example, since the prepaid service is used
more often by young people, our findings suggest that the relations
among young people are more biased. Obviously hypotheses like this one
need more research until we can say anything conclusive.

Data sets including communication through several different channels,
such as email, online chatting and real life conversations, would be
especially well suited for studying reciprocity, but unfortunately
such data sets are difficult to acquire in the large scale
required. The methods used here can however be readily applied to any
such data, as long as the data includes information on who initiated
the communication.

\section*{Acknowledgment}

This work is partially supported by the Academy of Finland, the
Finnish Center of Excellence programme 2006--2011, proj. 129670. We
would also like to thank prof. Albert-L\'{a}szl\'{o} Barab\'{a}si of
Northeastern University for providing us access to the unique mobile
phone data set and continuing collaboration thereof.


\section*{Appendix}
\label{appendix}

We constructed a reference network to find out how what portion of the
high edge bias values can be explained by the out-strength
distribution. Unlike in-strength, the out-strength can be seen to
reflect the activity of each user, and is closely controlled by each
individual. Thus is it feasible to think that each call made could
have been directed to some other contact instead of the one actually
called. We calculate the reference for the whole network, comprising
both the prepaid and postpaid users, and only separate the users
groups in the final reference network. This is based on the assumption
that customers do not know or care about the user type (prepaid or
postpaid) of the recipient of the call, and thus any call made to a
user of one type could have been made to a user of the other type.

There are obviously many ways of pursuing the goal of a bias
distribution centered around $b_{ij} = 0.5$. We decided to base our
method on the reasonable hypothesis about binomial distribution of
edge weights which was discussed in Section \ref{sec:results}: we
attempt to maximize the likelihood that the edge bias values come from
binomial distribution with $p=0.5$, that is, we try to find a new set
of edge weights $\mathbf{w} = \{ w_{ij} \}_{e_{ij}\in\set{E}}$ such
that
\[ \begin{split} 
\mathbf{w} = \mathrm{arg} \max & \, \Pi_{(i,j)\in
  \set{E}} \textrm{Pr}(w_{ij}\,|\,w_{ij} \sim \mathrm{Bin}(w_{ij}+w_{ji},\,0.5)) \\
\mathrm{s.t.} & \sum_{j}w_{ij} = s_i~\forall v_i \in V,
\end{split} \]
where $s_i$ is the out-strength of node $v_i$ in the original
network. The topology of the original network is also preserved, which
means that the edge set $\set{E}$ remains the same during the
optimization.

Taking a logarithm and writing the binomial probabilities explicitly,
we get
\[ \begin{split}
\mathbf{w} & = \mathrm{arg} \max \sum_{e_{ij}\in \set{E}} \log
\frac{(w_{ij}+w_{ji})!}{w_{ij}! w_{ji}!}p^{w_{ij}}(1-p)^{w_{ji}} \\
& = \mathrm{arg} \max f(\mathbf{w}),
\end{split} \]
where $f(\mathbf{w})$ is the global target function
\begin{equation} \label{eq:fw}
f(\mathbf{w}) = \sum_{e_{ij}\in \set{E}} \Bigl(
\sum_{u=1}^{w_{ij}+w_{ji}} \log u -
\sum_{u=1}^{w_{ij}} \log u - \sum_{u=1}^{w_{ji}} \log u \Bigr).
\end{equation}

\begin{figure} \begin{center}
    \includegraphics[width=10cm]{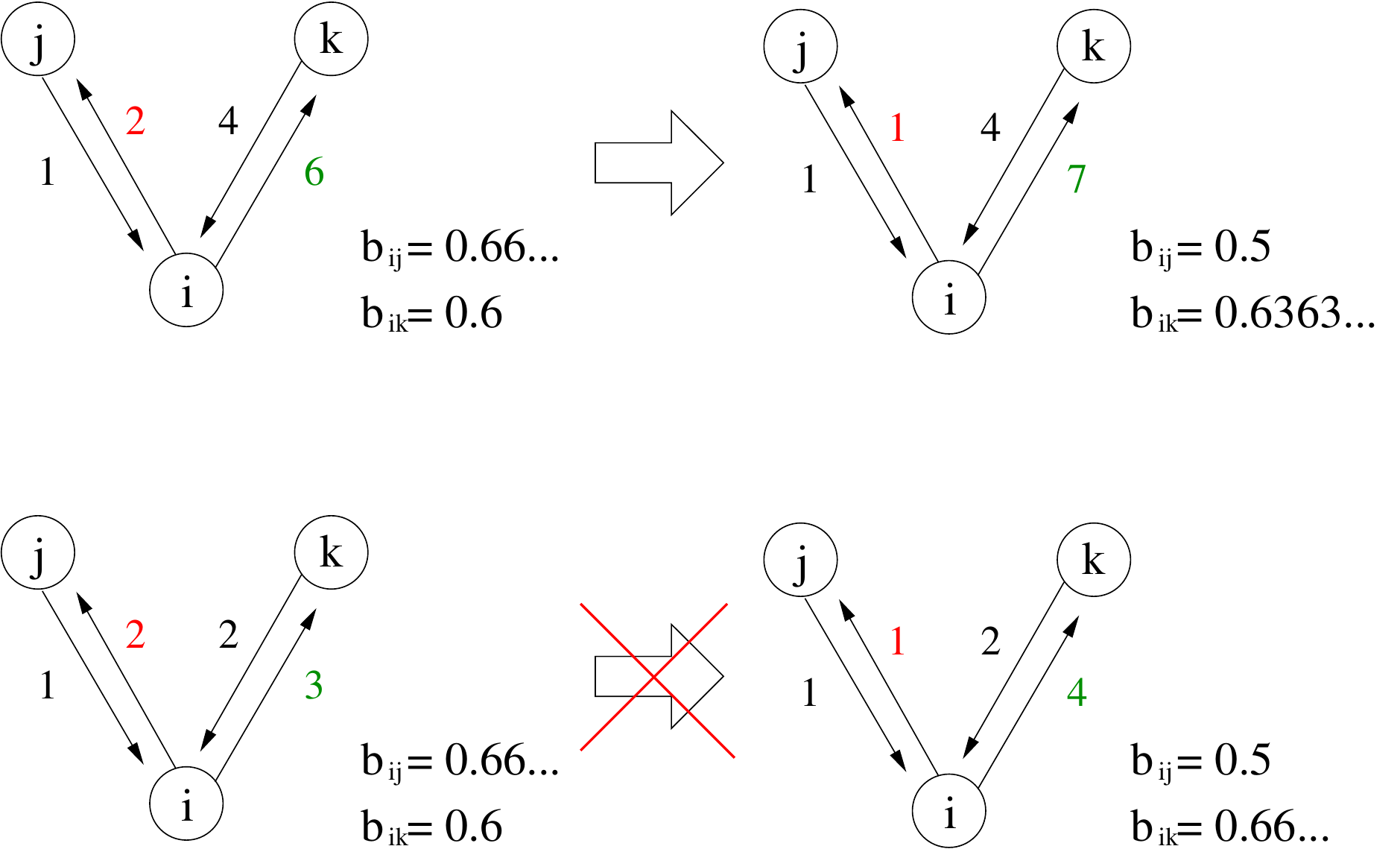}
    \caption{\textbf{(Top)} One unit of weight is moved from edge
      $(v_i,\,v_j)$ to $(v_i,\,v_k)$. Since $b_{ij}(0) =
      0.6\overline{6} > 0.\overline{63} = b_{ik}(1)$, this change
      increases the value of the target function. This means that the
      relative increase in the likelihood of the edge $(v_i,\,v_j)$
      (33.3 \%, from ${3 \choose 2}0.5^3 = 0.375$ to ${2 \choose
        1}0.5^2 = 0.5$) outweighs the decrease in the likelihood of
      the edge $(v_i,\,v_k)$ (21.4 \%, from ${10 \choose 6}0.5^{10}
      \approx 0.205$ to ${11 \choose 7}0.5^{11} \approx 0.161$):
      $(1+0.3\overline{3})(1-0.214) > 1$. \textbf{(Bottom)} Because
      $b_{ij}(0) = 0.6\overline{6} = b_{ik}(1)$, this move does not
      increase the value of the target function and is therefore not
      performed. In fact, since the two bias values turn out to be
      equal, this means that the increase in the likelihood of the
      edge $(v_i,\,v_j)$ (33.3 \%) exactly matches the decrease in the
      likelihood of the edge $(v_i,\,v_k)$ (25 \%, from ${5 \choose
        3}0.5^{5} = 0.3125$ to ${6 \choose 4}0.5^{6} \approx 0.234$):
      $(1+0.3\overline{3})(1-0.25) = 1$.}
    \label{fig:weight_change}
\end{center} \end{figure}

Finding the global optimum of this function is not straightforward, as
it is likely to have several local optima. We carry out the
optimization in two-phases. The first phase is simulated annealing,
which aims at avoiding bad local optima. One step of the simulated
annealing algorithm selects a random node $v_i$ and moves one unit of
weight\footnote{Since the edge weights are in this case call counts,
  they are discrete and reasonably small.} from edge $(v_i,\,v_j)$ to
$(v_i,\,v_k)$, where $v_j$ and $v_k$ are random neighbors of $v_i$,
with a probability proportional to the change of the target function
$f(\mathbf{w})$, given by
\[ \begin{split}
\Delta f(\mathbf{w}) & = f(\mathbf{w};\, w_{ij}
\mapsto w_{ij} - 1,\,w_{ik}
\mapsto w_{ik} + 1) \\
& ~~~ - f(\mathbf{w}) \\
& = \log \frac{w_{ij}(w_{ik}+w_{ki}+1)}{(w_{ij}+w_{ji})(w_{ik}+1)} = \log \frac{b_{ij}(0)}{b_{ik}(1)}.
\end{split} \]
In the last step we have defined
\[
b_{ij}(\delta) = \frac{w_{ij} + \delta}{w_{ij} + w_{ji} + \delta}
\]
which is the bias of edge $(v_i,\,v_j)$ after adding weight $\delta$. This
final result is surprisingly simple: the target function value is
increased whenever $b_{ij}(0) > b_{ik}(1)$. Interestingly, while
$f(\mathbf{w})$ was derived by maximizing the likelihood of binomial
distributions, we end up comparing the edge biases. The two biases
involved in the comparison implicitly account for the form of the
binomial distribution; see Figure \ref{fig:weight_change} for
further clarification.

When the temperature of the simulated annealing becomes low, there is
little or no randomness left, but the algorithm is still quite slow.
We therefore use a greedy algorithm as the second phase of the
optimization. The greedy algorithm starts from where the simulated
annealing finished, and simply descends to the local optimum as
efficiently as possible.

The greedy optimization goes through the nodes in a round-robin
manner, but unlike simulated annealing, improves the target function
by changing weights around one node until no further progress is
possible. It works by selecting the two edges that give the maximum
improvement of the target function, $j^* = \arg_j\max\{b_{ij}(0)\}$
and $k^* = \arg_k\min\{b_{ik}(1)|k\neq j^*\}$, and changing $\delta^* =
\left\lceil \frac{w_{ij}w_{ki}-w_{ji}w_{ik}-w_{ji}}{w_{ji}+w_{ki}}
\right\rceil$ units of weight from edge $(v_i,\,v_{j^*})$ to
$(v_i,\,v_{k^*})$. $\delta^*$ is the largest amount of weight that still
improves the target function, defined as the largest integer for which
$b_{ij}(1-\delta) > b_{ik}(-\delta)$ is true. The optimum is reached when no
change occurs during one round through all nodes.

\end{document}